\documentclass[aps,prl,twocolumn,showpacs]{revtex4}
\usepackage{epsfig}
\usepackage{graphicx}

\begin{document}

\DeclareGraphicsExtensions{.eps,.EPS}

\title{Simultaneous Magneto-Optical Trapping of Bosonic and Fermionic Chromium Atoms}
\author{R. Chicireanu, A. Pouderous, R. Barb\'e, B. Laburthe-Tolra, E. Mar\'echal, L. Vernac, J.-C. Keller, and O. Gorceix}
\affiliation{Laboratoire de Physique des Lasers, UMR 7538 CNRS,
Universit\'e de Paris Nord, 99 Avenue J.-B. Cl\'ement, 93430
Villetaneuse, France}

\begin{abstract}
We report on magneto-optical trapping of fermionic $^{53}$Cr atoms.
A Zeeman-slowed atomic beam provides loading rates up to
3$\times10^{6}s^{-1}$. We present systematic characterization of the
magneto-optical trap (MOT). We obtain up to 5$\times10^{5}$ atoms in
the steady state MOT. The atoms radiatively decay from the excited P
state into metastable D states, and, due to the large dipolar
magnetic moment of chromium atoms in these states, they can remain
magnetically trapped in the quadrupole field gradient of the MOT. We
study the accumulation of metastable $^{53}$Cr atoms into this
magnetic trap. We also report on the first simultaneous
magneto-optical trapping of bosonic $^{52}$Cr and fermionic
$^{53}$Cr atoms. Finally, we characterize the light assisted
collision losses in this Bose-Fermi cold mixture.
\end{abstract}

\pacs{32.80.Pj, 39.25.+k, 42.50.Vk}
\date{\today}
\maketitle

\section{Introduction}

Although laser cooling generally assumes the presence of an almost
closed transition \cite{metcalf}, it is now possible to laser cool
in a magneto-optical trap (MOT) atomic species whose internal level
structure is more complex than that of alkali atoms. A more complex
electronic structure may lead to qualitatively new physics. In the
case of $^{52}$Cr, for example, for which Bose-Einstein condensation
was reached recently in the group of T. Pfau in Stuttgart, Germany
\cite{BECChrome}, anisotropic, long-range dipole-dipole interactions
play a visible role in the expansion of the BEC after it is released
from its trap \cite{Pfaudipole} - contrary to all previous
experimental data on BECs, for which all interaction effects could
be modeled by short-range isotropic interactions. In addition, the
optical manipulation of new atoms may find technical applications,
which is the case for chromium (Cr) \cite{chrome}, but also for
example, for group III atoms (such as Al, Ga, In) and for Fe
\cite{rehse}.

The optical manipulation of species with complex electronic
structure can nevertheless be a technical challenge when the cooling
transition is not closed, since more and more repumping lasers are then
needed. For example, laser cooling of molecules has never been investigated
because of the many vibrational levels associated with a given electronic
transition. Here, we report on the magneto-optical trapping of fermionic $^{53}$%
Cr atoms. As with bosonic $^{52}$Cr atoms, the cooling transition is
not perfectly closed and atoms are depumped towards metastable dark
states, so that two repumpers are required to effectively close this
transition. In addition, and in contrast to $^{52}$Cr which has no
hyperfine structure, $^{53}$Cr nuclear magnetic moment is $I=3/2$,
and its electronic ground state has four hyperfine components.
Further repumpers are therefore needed, to prevent optical pumping
towards dark hyperfine states.

Obtaining a MOT for $^{53}$Cr atoms is appealing for several
reasons. Starting from a laser-cooled atomic vapor, it has been
possible to reach quantum degeneracy for nine bosonic atomic
species, which is one of the major achievements of atomic physics in
the last ten years. However, up to now only two fermionic species,
$^{40}$K and $^6$Li, were cooled down to quantum degeneracy
\cite{degenerate}. For both species, it was possible to reach the
strongly interacting regime by use of Feshbach resonances, and to
study the cross-over between the BCS and the BEC regimes. It is
therefore interesting to add a new species in the ultra-cold
fermionic playground. The main specificity of chromium is its large
magnetic moment of 6 $\mu_B$: in polarized degenerate Fermi seas,
dipole-dipole interactions will be the leading interaction term,
which is expected to produce new quantum phases
\cite{dipolarfermisea}. In addition, the atomic ground state
hyperfine structure together with its strong magnetic dipole moment
should lead to a rich Feshbach resonances spectrum. Finally we
report in this paper that it is possible to simultaneously trap
$^{53}$Cr and $^{52}$Cr atoms, and discuss the perspectives to
produce chromium Bose-Fermi mixtures in the quantum regime.

\section{Experimental Setup}

\subsection{Laser system}

\begin{figure}[h]
\centering
\includegraphics[width=3.5in]{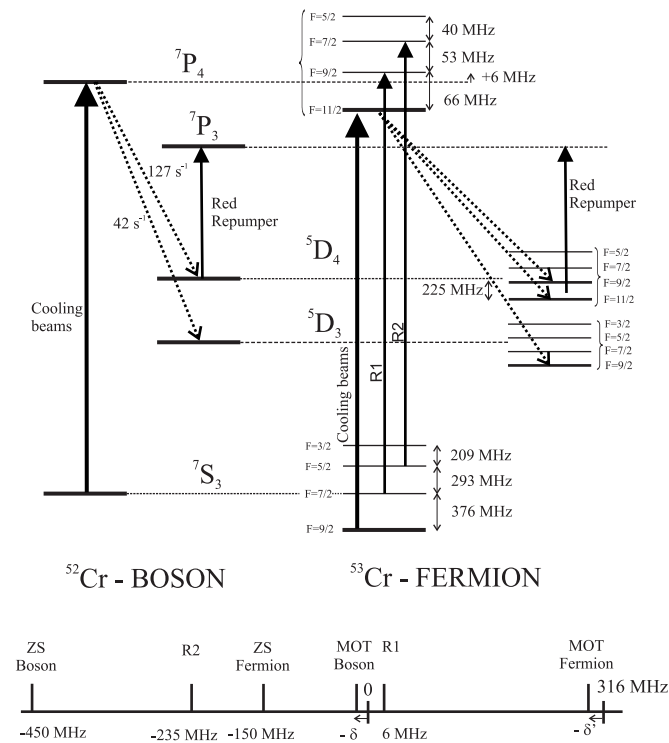}
\caption{\setlength{\baselineskip}{6pt} {\protect\scriptsize
Relevant energy levels for cooling and trapping $^{52}$Cr and
$^{53}$Cr including the leaks to metastable states and (bottom)
sketch of the laser frequencies used in our experiment ($\delta $
and $\delta ^{\prime }$ stand for the MOT detunings). Transition
probabilities to the bosonic metastable states are from
\cite{stuhler}, hyperfine structure splitting of the fermionic
isotope are from \cite{hyperfine}. We deduced from our experiment
the isotopic shift between the $^{52}$Cr $^7$S$_3 \rightarrow
^7$P$_4$ transition and the $^{53}$Cr $F=7/2 \rightarrow F'=9/2$
transition to be  $6 \pm 1$ MHz.}} \label{schemaNiveaux}
\end{figure}

Figure 1 shows a simplified energy-level diagram for $^{52}$Cr and $^{53}$%
Cr atoms. The $^7$S$_3\rightarrow^7$P$_4$ transition ( $\lambda$ =
425.51 nm, natural linewidth $\gamma  / 2 \pi = $ 5.02 MHz,
saturation intensity $I_s$ = 8.52 mW.cm$^{-2}$ ) is used both to
decelerate the two isotopes in a Zeeman slower, and to cool and trap
them in a MOT. Since no suitable diode laser source is yet available
at 425 nm, we generate 350 mW of blue laser light by frequency
doubling high-power light at 851 nm using a non-linear 8 mm long LBO
crystal in a resonant optical cavity. The light at 851 nm is
produced with an argon-ion laser (10.5 W) pumping a commercial
Ti:Sapphire laser from Tekhnoscan. The infrared power output is
typically 1.3 W. The frequency of the Ti:Sa laser is stabilized by
locking it to a Fabry-Perot (FP) reference cavity (whose finesse is
100). The FP spacer is made of Invar steel. The cavity is housed in
an evacuated chamber, which is isolated acoustically and thermally.
The length of the Fabry Perot cavity is locked to the
$^7$S$_3\rightarrow ^7$P$_4$ transition of $^{52}$Cr by use of a
standard saturation spectroscopy setup on an hollow cathode lamp.
The doubling cavity is kept in resonance with the infra-red light
using a H\"{a}nsch-Couillaud locking technique \cite{Hansch}. We
estimate the blue laser jitter to be around one MHz. Several
Acousto-Optic Modulators (AOMs) are necessary to generate the laser
frequencies needed for slowing and trapping both chromium isotopes
(see Fig \ref{schemaNiveaux}).

The $^7$S$_3\rightarrow^7$P$_4$ transition is slightly leaky since
excited atoms can decay from $^7$P$_4$ to dark $^5$D states (see Fig
\ref{schemaNiveaux} and \cite{MOTCr1}). To compensate for that
effect, repumping lasers can be used to pump atoms from the
metastable D states back to the ground $^7$S$_3$ state via the
$^7$P$_3$ state. Their wavelength are in the range of $660$ nm and
are thus referred as the 'red' repumpers in the following.  In our
setup they are produced by external cavity laser diodes, each locked
to the Fabry-Perot reference cavity using a Pound-Drever-Hall scheme \cite{pdh}%
.

\subsection{The Chromium Source}

Our chromium oven is a commercial Addon high temperature effusion
cell. The realization of a robust chromium source is a technical
challenge. Indeed, chromium is a very refractory metal, with a
melting point of 1857 Celsius. Furthermore, Cr tends to react with
most of the materials used to construct high temperature ovens. For
example, direct contact with Tantalum (Ta) or Tungsten (W) must be
avoided because Cr forms alloys with low temperature melting with
them.

In our setup, the cell is resistively heated with two self
supporting W filaments. The main crucible is in Ta. A chromium bar
of 20 g (99.7 percent pure) is enclosed in a second crucible made of
ultra pure alumina, which is inserted inside the Ta crucible. The
temperature is measured with a W/Re thermocouple. A water cooled
thermal shield maintains the vacuum chamber at a moderate
temperature of 40 Celsius. The effusive source is terminated by a 2
mm diameter aperture. A 4 mm diameter aperture, set at a distance of
5 cm from the emission point, defines the direction of the atomic
beam. The cell works in the horizontal plane and is connected to the
chamber through a CF40 orientable flange. This provides the
necessary fine tuning for the orientation of the atomic beam in
order to align the thermal beam along the axis of the one meter long
Zeeman slower tube. The atomic beam can be switched off and on
within 200 ms with a mechanical shutter. The oven chamber is pumped
with a 250 L.s$^{-1}$ turbo-pump. We usually work with an oven at a
temperature of 1500 Celsius, and the pressure in the oven chamber is
then equal to $3\times10^{-10}$ mbar.

\subsection{The Zeeman Slower}

A Zeeman slower (ZS) connects the oven chamber to the MOT chamber.
It is made of three sets of copper coils wrapped around a
one-meter-long steel tube, which has an internal diameter of $1.4$
cm, and is connected to the oven chamber through a $10$ cm-long
flexible hose. At the entrance of the ZS tube, a $25$ cm-long
differential pumping tube having a $0.9$ cm internal diameter allows
limiting pressures in the experimental chamber below
$5\times10^{-11}$ mbar when the oven is on.

\begin{figure}[h] \centering
\includegraphics[width=3.5in]{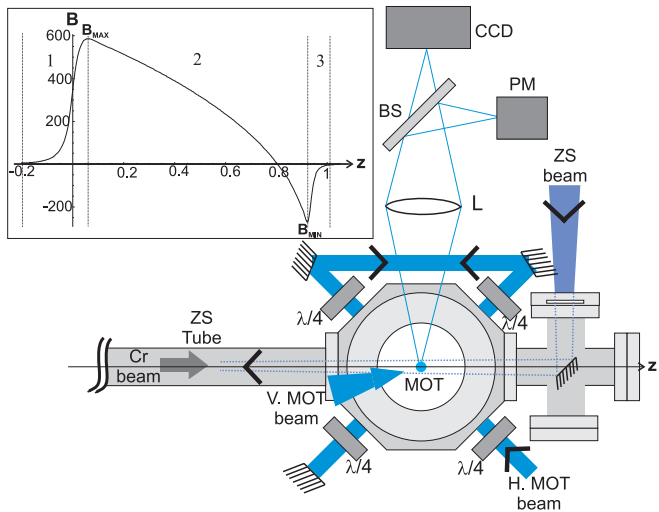}

\caption{\setlength{\baselineskip}{6pt} {\scriptsize  Schematic of
the experimental chamber showing the MOT beams, the coupling of the
ZS beam, and the MOT fluorescence detection. $L$, lens; $\lambda
/4$, quarter waveplate; BS, beam splitter; V., vertical; H.,
horizontal. The insert shows a plot of the experimental ZS magnetic
field in G as a function of the z position along its axis in m. See
text for a discussion of the three zones.}} \label{PlanManip}

\end{figure}

The experimental magnetic field profile achieved is a usual
positive- to- negative
square root profile, allowing one to get a well-defined final velocity (see \cite%
{Metcalf} for a discussion on the extraction at the end of a ZS, and
\cite{Slowe} for a recent and detailed paper on a ZS). It consists
in three parts: the "branching zone" at the beginning of the tube
where the magnetic field rises and quickly reaches its maximum
intensity B$_{max}
> 0$; the "slowing zone" with a smooth decrease towards B$_{min} <
0$; and a "final decrease zone" from B$_{min}$ to zero (these three
zones are shown in the insert of Fig \ref{PlanManip}). The small
diameter of the final coil makes it possible to obtain a short
"final decrease zone". This allows us to produce the MOTs close to
the exit of the tube ($10$ cm), which reduces losses due to
transverse expansion at the end of the slower. The experimental
magnetic B field values are B$_{max} = 460$ G and B$_{min} = -260$
G, which, together with the detuning of the ZS laser beam ($-450$
MHz), define a capture velocity $V_{c}$ of $460$ m.s$^{-1}$, and a
final velocity of $40$ m.s$^{-1}$.

At the exit of the oven, the populations of the fermionic isotope
(respectively bosonic) are equally distributed into the $28$ ($7$)
Zeeman sublevels of the four (one) hyperfine ground states (ground
state). Only the atoms reaching the "slowing zone" in the
$\left|F=9/2,m_{F}=9/2 \right\rangle$ ($m_{J}=3$) Zeeman sublevel
can be slowed down in the ZS. In order to get a theoretical estimate
of the fluxes of atoms slowed in the ZS, we carried out optical
pumping calculations for both the bosonic and fermionic isotopes.
While the bosonic case is easy to deal with, the fermionic one is
much more complicated. First the number of levels ($28$ in the
$^{7}S_{3}$ and $36$ in the $^{7}P_{4}$ ) is large. In addition, the
Zeeman energy shifts become comparable with the hyperfine structure
for low values of the B field (a few Gauss for the excited
$^{7}P_{4}$ state), so that the calculations must involve the true
eigenstates, which are neither the hyperfine $(F,m_{F})$ states nor
the Zeeman $(m_{J},m_{I})$ states.

We found that most of the $^{52}Cr$ atoms whose velocity is smaller
than or equal to $V_{c}$ are optically pumped into the $m_{J}=+3$
state by the ZS slowing beam inside the "branching zone". On the
contrary, no substantial accumulation in the $\left| F=9/2,m_{F}=9/2
\right\rangle$ state is expected for the $^{53}Cr$ atoms as they
travel through the "branching zone" ($6\%$ instead of $3.5\%=1/28$).
Taking into account both this different behavior with respect to
optical pumping and the isotopic proportion in natural chromium, we
expect the ratio of the flux of slowed $^{53}Cr$ atoms to the flux
of slowed $^{52}Cr$ atoms to be equal to $0.06\times 9.5/84=0.7\%$.
The experimental ratio, deduced from the analysis of MOT loading
sequences, is larger, almost equal to $2\%$ at 1500 Celsius. It
should be noted though that less laser power is available for the
Boson manipulation than for the Fermion in our experiment.

In the ZS, the $^{53}Cr$ atoms experience a bad crossing pattern:
for B$=25$ G, the two excited eigenstates adiabatically connected
respectively to $\left|F^{\prime }=11/2,m_{F^{\prime
}}=11/2\right\rangle$ and to $\left|F^{\prime }=9/2,m_{F^{\prime
}}=7/2\right\rangle$ at B$=0$ G are degenerate, so that any $\sigma
^{-}$ polarized ZS laser depumps the atoms towards the $F=7/2$
ground state. In our experiment we clearly do not expect the ZS
laser beam to be perfectly $\sigma ^{+}$ polarized: we use an
undervacuum metallic Al mirror to reflect the ZS beam and align it
along the chromium beam (see Fig \ref{PlanManip}); this mirror is
coated by chromium as time goes on, and its reflectivity properties
slowly change in time \cite{miroirvide}. Optical pumping
calculations show that about $20\%$ of the atoms can be lost if
$10\%$ of the ZS power is $\sigma ^{-}$ polarized. To repump
$^{53}$Cr atoms in the ZS, we use the $^{52}$Cr ZS beam, which is
(at 25 G) resonant with the transition between the states
adiabatically connected to  $\left|F=7/2,m_{F}=7/2\right\rangle$ and
$\left|F^{\prime }=9/2,m_{F^{\prime }}=9/2\right\rangle$ (at B=0 G).
Experimentally we got the optimal $^{53}$Cr
 MOT loading rates with a power of $3$ mW in the $^{52}$Cr ZS beam, and $
30$ mW in the $^{53}$Cr ZS beam. The $1/e^2$ radius of the $^{52}$Cr ZS beam and of the $^{53}$Cr ZS beam is respectively 3 mm and 3.6 mm at the MOT position, and both beams are approximately focused at the oven position. The presence of the $^{52}$Cr ZS beam as a repumper in the fermionic Zeeman slower results in an increase of about $40\%$ in the final number of atoms in the $^{53}$Cr MOT.

We also performed a standard 1D transverse cooling in the oven
chamber along the horizontal direction, using a laser having the
same frequency as the MOTs lasers. In order to compensate for the
red detuning of these lasers we applied a B field of a few Gauss
along the transverse cooling beam direction. In the optimal
experimental configuration, the transverse cooling laser has a power
of 10 mW, and induces a gain of $2$ to $2.5$ in the steady state MOT
populations. Finally we obtained MOT loading rates of the order of
$1.6\times10^{8}$ $(3\times10^{6})$ atoms.s$^{-1}$ for $^{52}Cr$
$(^{53}Cr)$ at $T=1500$ $C$.

\subsection{The Experimental Chamber}

Our experimental chamber is a compact octogonal chamber with eight
CF40 viewports in the horizontal plane, and two CF100 viewports for
the coupling of vertical laser beams (see Fig \ref{PlanManip}). A
150 L.s$^{-1}$ ion pump and a Ti-sublimation pump maintain the
pressure at about $5\times10^{-11}$ mbar. Two horizontal external
coils in anti-Helmholtz configuration create a quadrupole field at
the MOT position. The MOT laser beam setup uses two independent
arms: a vertical beam is retroreflected, and a single horizontal
retroreflected beam provides all four horizontal cooling beams (as
shown in Fig \ref{PlanManip}). The atomic clouds are imaged both on
a PhotoMultiplicator (PM), and a CCD Camera with a $12.5$ cm focal
length lens.

\section{Experimental Results}

\subsection{Magneto-optical trapping of $^{52}$Cr atoms}

As a first step, we report the achievement of a magneto-optical trap
with bosonic $^{52}$Cr atoms. After transverse optical cooling, the
atoms are slowed in the ZS, and trapped in the MOT. Because
$^{52}$Cr has no hyperfine structure, a single frequency is
sufficient to slow the atoms in the ZS, and another to cool and trap
atoms in a MOT. The vertical magnetic field gradient is 18
G.cm$^{-1}$, the MOT beams are typically detuned by a few linewidths
below the atomic resonance, and their $1/e^2$ radius is 7 mm.

A usual calibration (see for example \cite{MOTCr2}) of the MOT
fluorescence measured on the PM, which takes into account the
collection solid angle of the imaging lens and the 3/7 average
squared Clebsch-Gordan coefficient of the trapping transition,
allowed us to estimate the following numbers. Our loading rate is up
to $1.6\times10^{8}$ atoms per second, at an oven temperature of
1500 Celsius. We measured a temperature of $100\pm20$ $\mu K$ from
the free expansion of the atomic cloud released from the MOT. We
obtained a maximum atom number of $(4\pm.2\pm1)\times10^{6}$
\cite{uncertainty}, corresponding to a peak atom density of
8$\times10^{10}$ cm$^{-3}$ (with a 30$\%$ systematic uncertainty).

These results mainly reproduce similar measurements made in the
groups of J. McClelland at NIST in Gaithersburg, Ma, USA, and T.
Pfau (\cite{MOTCr1},\cite{MOTCr2}). First, we observed a strong
one-body loss, due to the fact that the cooling transition is not
perfectly closed. Atoms in the $^7P_4$ state can undergo an
intercombination transition, which leaves them in either the $^5D_4$
or the $^5D_3$ states. These metastable dark states are insensitive
to the MOT light, but can remain trapped in the MOT B field
gradient. We checked that atoms in these two states can be repumped
(via the $^7$P$_3$ state), using either one of two single mode
extended cavity laser diodes running at 653.97 nm and 663.18 nm,
which leads to a slight increase in the total number of atoms in the
steady state MOT.

The second striking feature of $^{52}$Cr MOTs is the strong two-body
inelastic loss rate, which was measured in \cite{MOTCr2}. We
reproduced the measurements reported in this reference, and found an
inelastic loss rate of $(6.25\pm.9\pm1.9)\times10^{-10}$
cm$^3$.s$^{-1}$, when the MOT beams are detuned by 10 MHz from
resonance, and have a total intensity of 116 mW.cm$^{-2}$. We
briefly describe the experimental procedure below.

\subsection{Magneto-optical Trapping of $^{53}$Cr Atoms}

The fermionic $^{53}$Cr isotope has already been cooled by
collisions with a cold buffer gas \cite{doyle}, but we are the first
group to report on properties of the magneto-optical trap of this
isotope, which is the first important new experimental result of
this paper \cite{Pfau}.

The optical manipulation of $^{53}$Cr is not a straightforward
extension of the work concerning $^{52}$Cr. The main difference
between these isotopes is the complex hyperfine structure of
$^{53}$Cr, whose electronic ground state has four different
hyperfine components. Although it may be expected that up to three
repumpers should be needed to operate the MOT, as well as the ZS,
the experiment shows that only two repumpers are needed in the MOT
(repumper $R1$ and $R2$ in Fig \ref{schemaNiveaux}), and only one in
the ZS. All these repumpers are derived from a single laser beam,
using AOMs. In addition to these 'blue' repumpers, three 'red'
repumpers are required to prevent atoms from ending up in the
metastable states $\left| ^5D_4, F=9/2,11/2 \right\rangle$ or
$\left| ^5D_3, F=9/2 \right\rangle$. Due to laser availability when
we performed the reported experiments, we were only able to repump
atoms from the $\left| ^5D_4 \right\rangle$ state.

We first achieved a MOT with $^{53}$Cr without any 'red' repumper.
Atoms exiting the oven are transversally cooled by a near resonant
$\left| ^7S_3, F=9/2 \right\rangle$ to $\left| ^7P_4, F=11/2
\right\rangle$ laser beam orthogonal to the atom beam, before being
decelerated in the ZS. When no transverse cooling is achieved before
the atoms enter the ZS, the MOT atom number is reduced by typically
60 percent. As explained earlier, a blue repumping laser detuned by
300 MHz from the slowing transition is mixed with the ZS laser beam
to prevent losses at the bad crossing point of the ZS. Typically a
few mW is enough for this repumper, and without it the number of
atoms in the MOT is only reduced by 30 percent.

The MOT 'blue' beams consist in three different frequencies. The
cooling beam is typically detuned by a few linewidths from the
atomic transition. Two 'blue' repumpers $R1$ and $R2$ respectively
correspond to the transition $\left| ^7S_3, F=7/2 \right\rangle$ to
$\left| ^7P_4, F'=9/2
\right\rangle$ and to the transition $\left| ^7S_3, F=5/2\right\rangle$ to $%
\left| ^7P_4, F'=7/2\right\rangle$ (see Fig \ref{schemaNiveaux}). No
MOT could be obtained without $R1$, whereas if $R2$ is removed, the
number of atoms in the MOT is only reduced by 30 percent. We
inferred from this observation that repumping on the transition
$\left| ^7S_3, F=3/2\right\rangle$ to $\left| ^7P_4,
F=5/2\right\rangle$ would not significantly increase the number of
atoms in the MOT. The power in $R1$ and $R2$ is respectively 10 mW
and 3 mW. The $1/e^2$ radius of the $^{53}$Cr MOT cooling beams is
4.5 mm, and the $1/e^2$ radius of $R2$ is 5.2 mm. As for $^{52}$Cr,
the MOT magnetic field gradient is 18 G.cm$^{-1}$, and a
fluorescence analysis (with an average squared Clebsch-Gordan
coefficient of 2/5 for the trapping transition) was used to get
quantitative data.

>From the depletion rate of the MOT, we estimated the transition
probability from $^7$P$_4$ to the metastable states $^5$D$_{4,3}$ to
be 280 s$^{-1}$ (with an uncertainty of $33\%$ mainly of statistical
origin) which is significantly higher than the corresponding number
for the boson (169 s$^{-1}$, see Fig \ref{schemaNiveaux}). We were
able to partially cancel these losses with the available 'red'
repumper (repumping from the $^5$D$_4$ state). The steady state atom
number in our MOT is multiplied by a factor of 2 in presence of this
repumper, and the repumping efficiency saturates for only 2 mW of
red light (with a $1/e^2$ radius of 2.9 mm). No significant increase
in the MOT fluorescence was observed when an other red beam, detuned
by the 225 MHz hyperfine gap of the $^5$D$_4$ state shown in Fig
\ref{schemaNiveaux}, was added.

\begin{figure}[h] \centering
\includegraphics[width=3.5in]{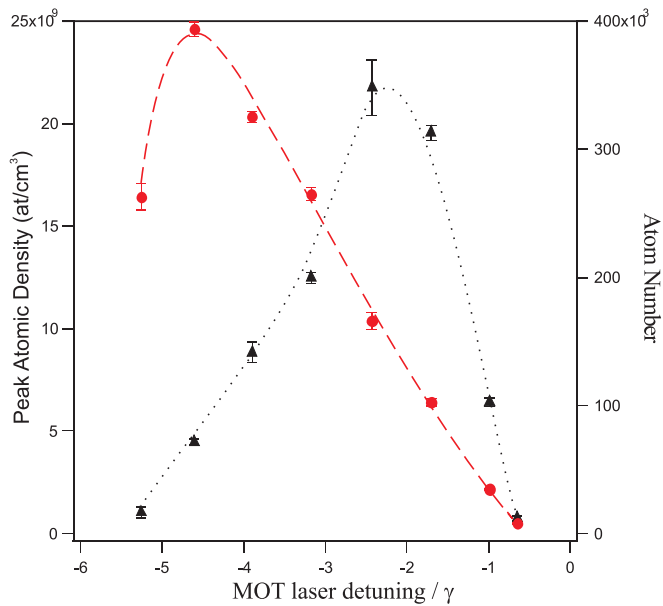}

\caption{\setlength{\baselineskip}{6pt}
{\scriptsize  Atom number (dots) and peak atom density (triangles) in the $^{53}$Cr MOT as a function of the normalized detuning of the MOT cooling beam compared to the $\left| ^7S_3, F=9/2\right\rangle$ to $%
\left| ^7P_4, F=11/2\right\rangle$ transition, with a total laser
intensity of the MOT beams equal to 200mW $cm^{-2}$. Error bars show
the dispersion over 4 data points, and the solid line is a guide for
the eye.}} \label{figdet}

\end{figure}

\begin{figure}[h] \centering
\includegraphics[width=3.5in]{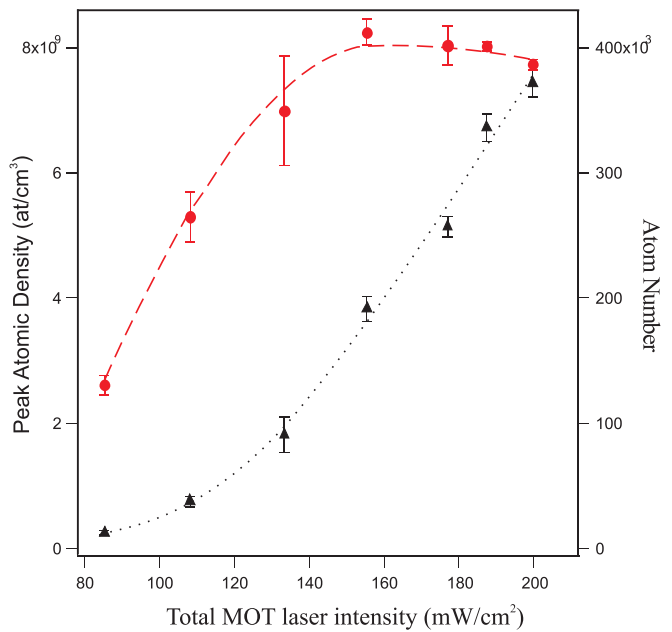}

\caption{\setlength{\baselineskip}{6pt} {\scriptsize  Atom number
(dots) and peak atom density (triangles) in the $^{53}$Cr MOT as a
function of the total power in the MOT cooling beams, for a detuning
of the MOT beams equal to $-21.5$ MHz. Error bars show the
dispersion over 4 data points, and the solid line is a guide for the
eye.}} \label{figint}

\end{figure}

We performed systematic studies to measure the total number of atoms
in the MOT in presence of the 'red' repumping laser, as a function
of the cooling laser intensity and detuning. Figure \ref{figdet}
shows that the MOT atom number and peak density reach a maximum for
different values of the MOT laser detunings. We observed up to
$(5\pm.25\pm1)\times10^5$ atoms in the $^{53}$Cr MOT. This steady
state number is significantly higher than what is expected from a
direct comparison with the boson case, if only the relative loading
rate (about 2$\%$) and loss rate of the two MOTs are taken into
consideration, showing the dramatic effect of inelastic collisions
for the denser bosonic MOT. Figure \ref{figint} demonstrates a
saturation of the MOT atom number with the MOT laser intensity,
while the peak density doesn't saturate for the laser power
available. The highest peak atomic density obtained was up to
2.5$\times$10$^{10}$ cm$^{-3}$ (with a systematic uncertainty of
30$\%$).

Even with one 'red' repumper on, the maximum number of $^{53}$Cr
atoms in the steady state MOT is below 10$^6$, which is most likely
not sufficient to reach quantum degeneracy via an evaporative
cooling procedure. As described below, a very large inelastic rate
coefficient in the $^{53}$Cr MOT, similar to the one measured for
$^{52}$Cr, limits the maximum number of atoms in the steady state
MOT. Therefore, we studied, similarly to the work reported in
\cite{accumulation} for the bosonic isotope, how fermionic
metastable atoms can be accumulated in the magnetic trap formed by
the quadrupole field of the MOT, before being repumped back into the
ground state. A similar accumulation process has been reported for
Strontium atoms in \cite{Nagel}, but requires a much higher value of
the B field gradient.

\subsection{Accumulation of $^{52}$Cr and $^{53}$Cr Atoms in a
Magnetic Quadrupole Trap}

Chromium offers a nice way to decouple cooling and trapping: as
stated above, atoms are slowly optically pumped into metastable
states, whose lifetimes are very large (and unknown), and whose
magnetic moments are large enough to be trapped by the MOT gradient
magnetic field, provided they happen to be produced in a low-field
seeking state. In our experiment, we first reproduced the results of
\cite{accumulation}, and used the same procedure to accumulate
metastable $^{53}$Cr atoms in the magnetic quadrupole field of the
MOT.

In the case of $^{52}$Cr, the number of trapped metastable atoms
saturates in a timescale which is fixed by inelastic collisions with
atoms in the excited $^7$P$_4$ state \cite{accumulation}. In the
case of $^{53}$Cr MOT, the steady state number of atoms in the
$^7$P$_4$ state is about ten times smaller, and we found that the
timescale for accumulating $^{53}$Cr metastable atoms in the
magnetic trap is set by collisions with fast atoms coming from the
oven. It is found to be on the order of 8 s when the oven is
operated at 1500 Celsius, which is equal to the measured lifetime of
the metastable fermions in the magnetic trap when the atom shutter
is not closed (and smaller than the background collision lifetime
when the shutter is closed).

By accumulating atoms in the metastable state $^7$D$_4$, and after
repumping them using a 10 ms long 'red' pulse, we measured up to
$(8.5\pm.4\pm2.1)\times10^5$ $^{53}$Cr atoms in the magnetically
trapped $^7$S$_3$ ground state. This number was obtained for a
detuning of the MOT beams equal to 12.5 MHz, and a total intensity
in the MOT beams equal to 200 mW.cm$^{-2}$, which corresponds almost
to a tenfold increase compared to the steady state MOT atom number
for these parameters. This measurement was achieved close to
resonance, and should not correspond to the optimal $^{53}$Cr atom
number accumulated in the metastable state (see Fig \ref{figdet}).
In addition, such a number was measured by using only one 'red'
repumper, and the use of an other 'red' repumper for the $^5$D$_3$
state should give us a reasonable starting point for further cold
collision experiments, Feshbach resonance studies, and for
evaporative cooling down to quantum degeneracy.

This larger number of $^{53}$Cr atoms allowed us to carefully study
density dependent losses for the $^{53}$Cr MOT. The two-body loss
rate parameter measured for $^{53}$Cr is very large, typically
ranging from $(1\pm.2\pm.3)\times10^{-9}$ to
$(8\pm.8\pm2.5)\times10^{-9}$ cm$^3$.s$^{-1}$ for detuning of the
MOT laser varying from -20 MHz to -3MHz, and a total MOT laser
intensity of 170 mW.cm$^{-2}$. These values are similar to the
inelastic loss rates obtained for $^{52}$Cr MOTs. To our knowledge,
these very high inelastic loss rates in chromium MOTs are not
understood up to now, and will be studied in a forthcoming
publication. Here, we note that they are only slightly smaller than
the Langevin rate (see \cite{julienne}).

\subsection{Magneto-optical Trapping of a Cold Bose-Fermi Mixture of $^{52}$%
Cr and $^{53}$Cr Atoms}

In this paper, we report on the first simultaneous magneto-optical
trapping of $^{52}$Cr and $^{53}$Cr atoms. To perform this
experiment, we make use of a fortuitous quasi coincidence between
two transitions: the cooling transition for $^{52}$Cr and the '$R1$'
transition for $^{53}$Cr. As a consequence, the cooling beam for
$^{52}$Cr is used in the $^{53}$Cr MOT as $R1$. Another consequence
is that the Zeeman repumper of $^{53}$Cr is also simultaneously used
as the ZS beam for $^{52}$Cr, so that we do not need a separate AOM
to perform the experiment.

\begin{figure}[h] \centering
\includegraphics[width=3.5in]{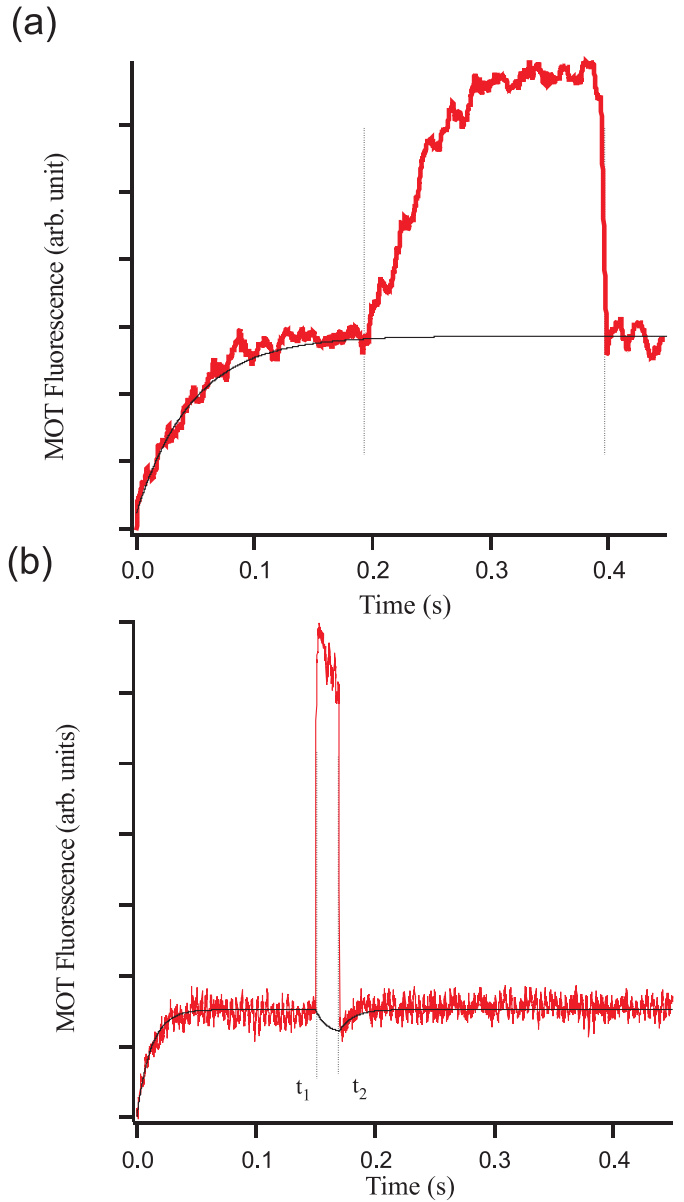}

\caption{\setlength{\baselineskip}{6pt} {\protect\scriptsize Loading
sequences for simultaneous Magneto-optical trapping of $^{52}$Cr and
$^{53}$Cr atoms. The MOT for $^{52}$Cr atoms is turned on at t=0,
whereas the AOMs for the $^{53}$Cr MOT are only turned on during the
time period indicated by the two vertical dashed lines. (a) Small
number of atoms: when the $^{53}$Cr MOT is turned off, the remaining
fluorescence signal of $^{52}$Cr atoms is identical to the one
before the $^{53}$Cr MOT was turned on, which means that no mutual
effects took place. The solid line is the result of a fit only
taking into account fluorescence of bosons. (b) Large number of
atoms: after $^{53}$Cr metastable atoms are repumped from the
magnetic trap and the large fermionic MOT is superimposed on the
$^{52}$Cr MOT, we observe a decrease in the total number of
$^{52}$Cr atoms. The solid line is the result of a fit, including
interspecies losses.}} \label{doubleMOT}
\end{figure}

Figure \ref{doubleMOT} (a) shows that we can load a $^{52}$Cr MOT,
and after it has reached its steady state number, superimpose on it
a $^{53}$Cr MOT by simply switching on the AOMs for the cooling
transition of the fermion, and $R2$. Both MOTs form approximately at
the same location according to our CCD fluorescence images. When the
$^{53}$Cr MOT laser beams are removed, the number of $^{52}$Cr atoms
immediately regains the same steady state value. This demonstrates
that neither these lasers nor the presence (in this regime) of
$^{53}$Cr cold atoms induce any substantial loss for $^{52}$Cr
atoms. However, the data in Fig \ref{doubleMOT} (a) correspond to
very few atoms (about 3$\times$10$^4$ $^{52}$Cr atoms, and
4$\times$10$^4$ $^{53}$Cr atoms).

In a separate set of experiments, we increased the number of
fermionic atoms, by first accumulating them in the trapped
metastable states during 10 s. We then loaded a $^{52}$Cr MOT,
repumped $^{53}$Cr atoms from the $^{5}D_4$ state, thus
superimposing about 7$\times$10$^5$ fermionic atoms on the bosonic
MOT, with an estimated peak density of fermionic atoms of about
4$\times$10$^{10}$ cm$^{-3}$. We then observed substantial loss of
$^{52}$Cr atoms, as shown in Fig \ref{doubleMOT} (b), which is a
proof that fermionic atoms provide a new loss channel for bosonic
atoms.

The corresponding loss rate is equal to $\beta_{BF}\int
n_{52}(r)n_{53}(r) d^3r $, where $\beta_{BF}$ is the interisotope
loss coefficient. If the number of fermionic atoms stays almost
constant, this rate, from the bosonic atoms point of view,
corresponds to a linear loss rate, being equal to $\beta_{BF}
\bar{n}_{53} N_{52}$, where $N_{52}$ is the number of bosonic atoms,
and $\bar{n}_{53}=N_{53} / \bar{V}_{53}$ is an average fermionic
density at the bosonic MOT position. $N_{53}$ is the number of
fermionic atoms and $\bar{V}_{53}$ is a volume depending on the
spatial distributions of the two isotopes. Writing the atomic
densities as $n_{i}(r)=n_{i0}f_{i}(r)$ gives $\bar{V}_{53}$ in terms
of integrals of the normalized density distributions $f_{i}(r)$
($i=52,53$, $n_{i0}$ is the peak density):
\begin{equation}
\bar{V}_{53}=\frac{\int f_{52}(r)d^3r\int f_{53}(r)d^3r}{\int
f_{52}(r)f_{53}(r)d^3r} \label{defVbar53}
\end{equation}
Experimentally, the MOT density distributions are Gaussian shaped.
With the CCD camera we measured the 1/$e$ horizontal and vertical
radii of the two MOTs ($w_{H52}=110$ $\mu$m, $w_{H53}=150$ $\mu$m,
$w_{V52}=110$ $\mu$m and $w_{V53}=140$ $\mu$m), and the small center
position separation in between them ($\Delta z=60$ $\mu$m, along the
vertical axis), with a precision of about $7\%$ for these numbers.
>From these values the numerical evaluation of the integrals in
(\ref{defVbar53}) is straightforward.

For the data in Fig \ref{doubleMOT}, the number of $^{52}$Cr atoms
is small enough that we can neglect light assisted collisions among
them, so that the evolution equation for the $^{52}$Cr MOT
population therefore reads:

\begin{equation}
\frac{dN_{52}}{dt}=\Gamma-\frac{N_{52}}{\tau}-\beta_{BF} \bar{n}_{53} N_{52}
\label{evol}
\end{equation}
where $\Gamma$ is the loading rate of the $^{52}$Cr MOT, $\tau$ is the depumping time towards metastable states,
 and the last term is only present when the fermionic MOT is applied (between $t_1$ and $t_2$ in Fig \ref{doubleMOT} (b)).
We fitted the loading sequence of the $^{52}$Cr MOT ($t<t_1$) to
deduce $\Gamma$ and $\tau$. We then adjusted $\beta_{BF}$ to
reproduce the reduction of $^{52}$Cr atoms when the $^{53}$Cr MOT is
removed at $t=t_2$. From this analysis (similar to the one performed
for a K-Rb MOT in \cite{marcassa}), we estimated the light assisted
inelastic loss coefficient between $^{52}$Cr and $^{53}$Cr to be
$\beta_{BF} = (1.8\pm.45\pm.65)\times10^{-9}$ cm$^3$.s$^{-1}$, for
$^{52}$Cr ($^{53}$Cr) MOT beams detuned by 10 (12.5) MHz, and having
a total intensity of 70 (200) mW.cm$^{-2}$. The solid line in Fig
\ref{doubleMOT} (b) is the result of our numerical fit.

$\beta_{BF}$ is on the same order of magnitude than the light
assisted loss coefficient that we measured both in a $^{52}$Cr MOT
and in a $^{53}$Cr MOT. However, physics of light assisted
collisions in the case of a mixture is different than for a
single-species MOT. For homonuclear molecules, the long range part
of the excited molecular potential asymptotically related to the
$(S,P)$ dissociation limit is dominated by resonant dipole-dipole
interaction, scaling as $C_3/R^3$. In contrast, for heteronuclear
molecules, the excited molecular potential is dominated by a
Van-der-Waals interaction scaling as $C_6/R^6$ (\cite{Grim}). Thus,
in a MOT composed of two different species, inter-species
light-assisted collisions typically occur at shorter interatomic
distances than intra-species light assisted collisions. In addition,
the $C_6$ coefficient for the molecular potentials related to the
($^{52}$Cr,$^7S_3$, $^{53}$Cr,$^7P_4$) dissociation limit is
positive. The associated molecular potentials are repulsive and
should not lead to light assisted losses in a MOT (as discussed
in\cite{Grim}). Therefore, we interpret losses related to light
assisted collisions in our double MOT as the consequence of a
process involving molecular potentials related to $^{52}$Cr,$^7P_4$
and $^{53}$Cr,$^7S_3$. However, because the shift between the
cooling transitions for $^{52}$Cr and for $^{53}$Cr is small, the
situation may be more complicated, because the molecular excited
potentials should be in $1/R^6$ at large distance, with a very large
$C_6$, and in $1/R^3$ at shorter distances. This situation is
similar to dipole-dipole interactions in cold Rydberg gases
\cite{Gallagher}.

A complication for continuously collecting large numbers of atoms in
our mixed-species MOT arises from the fact that the $^{52}$Cr MOT is
substantially altered by the presence of the $^{53}$Cr ZS beam. We
attribute this effect to the fact that slowed $^{52}$Cr atoms are
pushed by the $^{53}$Cr Zeeman slower beam as they approach the MOT
region (as the ZS magnetic field goes to zero, the bosonic atoms
slowly travel through a region where they are resonant with the
$^{53}$Cr ZS laser frequency). Our observations suggest that the
best strategy to obtain large samples of cold $^{53}$Cr and
$^{52}$Cr atoms is to sequentially accumulate them in the metastable
states, before repumping both isotopes in their electronic ground
state.

\section{Conclusion}
We have reported the magneto-optical trapping of cold fermionic
$^{53}$Cr atoms. Magnetic trapping of metastable atoms in the
quadrupole field of the MOT was demonstrated, yielding an increase
in the number of atoms by a factor of seven. These features are
encouraging as a first step in producing a degenerate dipolar fermi
gas. Furthermore we demonstrated the achievement of a dual-isotope
boson-fermion MOT. This opens the way to two isotope collision
studies, including the search for inter isotope Feshbach resonances,
and the obtention of sympathetic cooling. It also opens the way to
the potential realization of a quantum degenerate boson-fermion
mixture involving dipolar species.

\vspace{1cm}

Acknowledgements: LPL is Unit\'e Mixte (UMR 7538) of CNRS and
of Universit\'e Paris Nord. We acknowledge financial support from Conseil R\'{e}%
gional d'Ile-de-France (Contrat Sesame), Minist\`{e}re de
l'Education, de l'Enseignement Sup\'{e}rieur et de la Recherche, and
European Union (FEDER - Objectif 2). We also thank C. Chardonnet, V.
Lorent and H. Perrin for their help in starting this project, as
well as J. McClelland and T. Pfau for their many friendly advices.


\begin{thebibliography}{99}
\bibitem{miroirvide} A direct coupling through a viewport must be avoided because the atomic chromium beam
would rapidly coat it; J. McClelland and T. Pfau, personal
communications.
\end{thebibliography}
\end{document}